\documentstyle[prl,aps,twocolumn,epsf,floats]{revtex}

\begin{document}

\draft

\title{Power laws and collapsing dynamics of a trapped Bose-Einstein
condensate with attractive interactions}

\author{Hiroki Saito and Masahito Ueda}
\address{Department of Physics, Tokyo Institute of Technology,
Tokyo 152-8551, Japan \\
and CREST, Japan Science and Technology Corporation (JST), Saitama
332-0012, Japan}

\date{\today}

\maketitle

\begin{abstract}
The critical behavior of collective modes and the collapsing dynamics of
trapped Bose-Einstein condensates with attractive interactions are
studied analytically and numerically.
The time scales of these dynamics both below and above the critical point
of the collapse are found to obey power laws with a single parameter of
$N/N_c - 1$, where $N$ is the number of condensate atoms and $N_c$ is the
critical number.
The collapsing condensate eventually undergoes rapid implosion, which
occurs several times intermittently, and then the implosion turns to an
explosion.
The release energy of the explosion is found to be proportional to the
square of the interaction strength, inversely proportional to the
three-body recombination rate, and independent of the number of condensate 
atoms and the trap frequency.
\end{abstract}

\narrowtext

\section{Introduction}

Bose-Einstein condensates (BECs) of trapped atomic vapor have been
realized in several species.
The static and dynamical properties of BEC crucially depend on the sign of 
the interatomic interaction, which is, without an external field,
repulsive for ${}^{87}{\rm Rb}$~\cite{Anderson}, ${}^{23}{\rm
Na}$~\cite{Davis}, and ${}^{1}{\rm H}$~\cite{Fried}, and attractive for
${}^{7}{\rm Li}$~\cite{Bradley} and ${}^{85}{\rm Rb}$~\cite{Cornish}.
The interactions can be manipulated by applying the electric or magnetic
field, as the scattering length is sensitive to the external field near
the Feshbach resonance~\cite{Feshbach}.
Using this technique, one can control not only the strength of the
interaction~\cite{Inouye} but also its sign~\cite{Cornish}. 

In a spatially uniform three-dimensional system, the attractive
interaction between atoms makes BEC unstable.
In a spatially confined system, however, the zero-point energy serves as a
kinetic obstacle against collapse, allowing metastable BEC to be formed
if the number $N$ of BEC atoms is below a certain critical value
$N_c$~\cite{Ruprecht,Baym,Fetter,Dalfovo96,Dodd,Houbiers,Stoof,Ueda98,Kim,Wadati}.
Various properties of BEC below and above $N_c$ have been predicted.
Just below $N_c$, BEC may collapse via macroscopic quantum
tunneling~\cite{Kagan96,Shuryak,Stoof,Ueda98,Huepe}.
When $N$ exceeds $N_c$, BEC becomes unstable and exhibits complicated
collapsing dynamics.
Kagan {\it et al.}~\cite{Kagan98} have predicted that the implosion of BEC 
eventually turns to explosion due to loss of atoms by three-body molecular
recombination.
Because the lost atoms are replenished by a thermal cloud, the number of
BEC atoms again exceeds $N_c$, leading to the collapse-and-growth cycles
of BEC~\cite{Sackett97,Sackett98,Kagan98}.
Such dynamic behavior of collapsing BEC has been explored with recent Rice 
experiments~\cite{Sackett99}.
Ueda and Huang~\cite{UedaHuang} have shown that the collapse occurs
locally within a small region around the center of the trap, and this
result has been verified numerically~\cite{Elef}.
We have predicted~\cite{Saito} that implosions occur intermittently and in 
rapid sequence due to competition between a loss of atoms and the
attractive interactions.
When the interaction is suddenly switched from repulsive to attractive by
using the Feshbach resonance, various patterns such as a shell structure
may be formed in the atomic density~\cite{Saito}.

In recent experiments with ${}^{85}{\rm Rb}$ at JILA~\cite{Cornish},
nearly pure condensates have been produced in which the number of atoms
can be fixed, and both the strength and sign of the interaction can be
controlled by using the Feshbach resonance.
These results suggest that $N_c$ can be tuned with the number of BEC atoms
held fixed.
We may thus study the critical behavior of BEC by controlling the relevant 
parameters.

In the present paper, we study the dynamics of BEC with attractive
interactions at zero temperature both below and above the critical point.
We find that both collective-mode frequencies just below the critical
point and the time it takes BEC just above this critical point to collapse
obey power laws with a single parameter $N/N_c - 1$.
We use a time-dependent Gaussian trial wave function to analytically study
collective-mode frequencies and the collapsing dynamics, and compare the
obtained results with those obtained by numerically solving the Bogoliubov
equations and the time-dependent Gross-Pitaevskii (GP)
equation~\cite{GP}.
When the atomic loss due to inelastic collisions is taken into account,
the collapsing condensate undergoes intermittent implosions followed by an 
explosion.
We find the energy released by the explosion is proportional to $g^2 /
L_3$ and independent of the number of atoms and the trap frequency, where
$g$ is the strength of the interaction and $L_3$ is a coefficient of the
three-body recombination rate.

This paper is organized as follows.
Section \ref{s:2} formulates the problem and introduces an
effective-potential model based on a time-dependent Gaussian variational
wave function.
Section \ref{s:3} discusses the collective oscillations of BEC below
$N_c$.
Section~\ref{s:4} analyzes the collapsing dynamics of BEC above $N_c$.
In both regimes, our analytical and numerically exact results are in
excellent agreement.
Section~\ref{s:5} discusses the dynamics of implosion and explosion of BEC
by incorporating the atomic loss into our theory.
Finally, section \ref{s:conclusion} summarizes our results.

\section{Formulation of the Problem}
\label{s:2}

We consider a system of Bose-condensed atoms with mass $m$ and s-wave  
scattering length $a$, confined in an isotropic parabolic potential.
The Hamiltonian for the system is given by
\begin{eqnarray} \label{H}
\hat H & = & \int d{\bf r} \ \hat \psi^\dagger({\bf r})\! \biggl[
 -\frac{\hbar^2}{2m} \nabla^2 + \frac{m\omega_0^2{\bf r}^2}{2}
\nonumber \\
& & + \frac{2\pi\hbar^2 a}{m} \hat \psi^\dagger({\bf r}) \hat \psi({\bf
r})\biggr] \hat \psi({\bf r}),
\end{eqnarray}
where $\hat \psi({\bf r})$ is the field operator of the atoms.
The transition amplitude of the system from an initial $\psi_{\rm i}$ to a
final state $\psi_{\rm f}$ is expressed in terms of path integrals as
\begin{equation} \label{path}
\langle \psi_{\rm f}| e^{-\frac{i}{\hbar} \hat H (t_f - t_i)} | 
\psi_{\rm i}\rangle
= \int  {\cal D} \psi {\cal D} \psi^*
e^{\frac{i}{\hbar}S[\psi, \psi^*]},
\end{equation}
where the action $S[\psi, \psi^*]$ is given by
\begin{eqnarray} \label{S}
S[\psi, \psi^*] & = & N \hbar \int \!d{\bf r} \!\int\!dt 
\biggl[i\psi^*\frac{\partial}{\partial t} \psi 
+\frac{1}{2} \psi^* \nabla^2\psi - \frac{r^2}{2}\psi^*\psi
\nonumber \\
& & - \frac{g}{2} (\psi^* \psi)^2 \biggr].
\end{eqnarray}
Here the length, time, and $\psi$ are normalized in units of 
$d_0 = (\hbar / m\omega_0)^{1/2}$, $\omega_0^{-1}$ and 
$(N / d_0^3)^{1/2}$, respectively, and $g \equiv 4 \pi N a / d_0$.
The wave function is then normalized to unity.
Separating $\psi({\bf r},t)$ into amplitude and phase as
$\psi({\bf r},t)=A({\bf r},t)e^{i\phi({\bf r},t)}$, and substituting
this into Eq.~(\ref{S}), we obtain
\begin{eqnarray}\label{s1}
S & = & N \hbar \int \!d{\bf r} \!\int\!dt 
\biggl[ i\dot{A} A - A^2 \dot{\phi} + \frac{1}{2} A \nabla^2 A -
\frac{1}{2} A^2(\nabla \phi)^2
\nonumber \\
& & + \frac{i}{2} \nabla(A^2 \nabla \phi) - \frac{r^2}{2} A^2 -
\frac{g}{2} A^4 \biggr].
\end{eqnarray}
The requirement that the action be stationary
with respect to small variations in $\phi$ leads to 
\begin{equation}\label{cont}
\frac{\partial A^2}{\partial t}+\nabla(A^2\nabla\phi)=0.
\end{equation}
This is nothing but the equation of continuity that guarantees the
conservation of the number of atoms.
The stationary condition for small variations in $A$ is, together with
Eq.~(\ref{cont}), equivalent to the GP equation
\begin{equation} \label{GP}
i \frac{\partial}{\partial t} \psi = -\frac{1}{2} \nabla^2 \psi +
\frac{r^2}{2} \psi + g |\psi|^2 \psi.
\end{equation}

For the case of attractive interactions, it is reasonable to assume
that the amplitude takes a Gaussian form, with its width reduced due to
the attractive interaction.
The validity of this Gaussian approximation will be examined by comparison
with the numerically exact results in Secs.~\ref{s:3} and \ref{s:4}.
Because we want to study the dynamics of BEC, we allow the
amplitude $A({\bf r},t)$ to vary in time as~\cite{Baym,Stoof}
\begin{eqnarray} \label{A}
A({\bf r}, t) & = & \sqrt{\frac{1}{\pi^{3/2}d_x(t)d_y(t)d_z(t)}}
\nonumber \\
& & \times \exp\left[ -\frac{x^2}{2 d_x^2(t)}-\frac{y^2}{2
d_y^2(t)}-\frac{z^2}{2 d_z^2(t)} \right],
\end{eqnarray}
where $d_i(t)$ ($i=x,y,z$) is a time-dependent real variational parameter
characterizing the width of the wave function along the $i$-axis.
It can be shown that the equation of continuity (\ref{cont}) and the 
requirement that there should be no mass current both at the origin and 
at infinity uniquely determines the phase as
\begin{equation}\label{phase}
\phi({\bf r},t) = \frac{\dot{d}_x(t)}{2d_x(t)} x^2 +
\frac{\dot{d}_y(t)}{2d_y(t)} y^2 +\frac{\dot{d}_z(t)}{2d_z(t)} z^2.
\end{equation}
We thus obtain the variational wave function as~\cite{Garcia}
\begin{eqnarray} \label{Gtrial}
& & \psi_{\rm var}({\bf r}, t) =
\nonumber \\
& & \sqrt{\frac{1}{\pi^{3/2} d_x(t) d_y(t)
d_z(t)}} \exp\biggl\{ -\frac{x^2}{2 d_x^2(t)} [1 - i \dot{d}_x(t)
d_x(t)] \nonumber \\
& & - \frac{y^2}{2 d_y^2(t)} [1 - i \dot{d}_y(t) d_y(t)] - \frac{z^2}{2
d_z^2(t)} [1 - i \dot{d}_z(t) d_z(t)] \biggr\}.
\end{eqnarray}
The imaginary terms in the exponent in Eq.~(\ref{Gtrial}) describe the
mass current associated with the change in the width of the wave
function.
To find the most probable Feynman path within a functional space of the
complex Gaussian wave function~(\ref{Gtrial}), we substitute
Eq.~(\ref{Gtrial}) into Eq.~(\ref{S}), obtaining
\begin{eqnarray}\label{act2}
S & = & \frac{N \hbar}{4} \int dt \Biggl[ \sum_{i=x,y,z}
\left(\dot{d}_i^2(t) - \frac{1}{d_i^2(t)} - d_i^2(t) + \dot{d}_i(t)
\right) \nonumber \\
& & - \frac{\gamma}{d_x(t) d_y(t) d_z(t)} \Biggr],
\end{eqnarray}
where $\gamma \equiv \frac{4N}{\sqrt{2\pi}} \frac{a}{d_0}$.
The stationary condition $\delta S / \delta d_i = 0$ gives an equation of
motion for $d_i(t)$ as~\cite{Garcia}
\begin{equation} \label{eom}
\ddot{d}_i(t) =  -d_i(t) + d_i^{-3}(t) + \frac{\gamma}{2 d_x(t) d_y(t)
d_z(t) d_i(t)} = -\frac{\partial V_{\rm eff}}{\partial d_i},
\end{equation}
where
\begin{equation}\label{eff}
V_{\rm eff}\equiv\frac{1}{2}\sum_{i=x,y,z}\left(d_i^2+d_i^{-2}\right)
+\frac{\gamma}{2d_xd_yd_z}
\end{equation}
may be viewed as an effective potential for the widths of the wave
function $d_i(t)$~\cite{Stoof,Garcia}.

\section{Power-law behavior of collective modes}
\label{s:3}

It has been predicted in Ref.~\cite{Ueda98} that as the criticality is
reached, the monopole mode becomes softer according to the one-fourth
power $1 - N / N_c$.
In Ref.~\cite{Ueda98}, however, a sum rule is invoked in order to
determine the numerical coefficient.
Here we develop a general theory that determines both the power and the
numerical coefficient.

For an isotropic trapping potential, the stationary point of 
Eq.~(\ref{eom}), 
$d_x=d_y=d_z\equiv d_{\rm st}$, is determined by a positive root of
\begin{equation} \label{stat}
d_{\rm st}^5 - d_{\rm st} - \frac{\gamma}{2} = 0.
\end{equation}
For the case of attractive interaction $\gamma < 0$, this equation has a
positive root if the strength of the interaction $|\gamma|$ is smaller
than the critical value $\gamma_c$~\cite{Baym,Fetter},
\begin{equation} \label{stab2}
|\gamma| < \gamma_c \equiv \frac{8}{5^{5/4}},
\end{equation}
where $|\gamma| = \gamma_c$ corresponds to $d_{\rm st} = r_c \equiv 5^{-1/4}$.
The stationary point $d_{\rm st}$ is always greater than $r_c$ for $|\gamma| <
\gamma_c$.
The stability condition is given by
\begin{equation}\label{stab}
{\rm det} \left. \left( \frac{\partial^2V_{\rm eff}}{\partial d_i\partial
d_j} \right) \right|_{d = d_{\rm st}}
= (d_{\rm st}^{-4} - 5)(2d_{\rm st}^{-4} + 2)^2 < 0,
\end{equation}
where the derivative is evaluated at the stationary point $d_x = d_y = d_z
= d_{\rm st}$.
It follows from Eqs.~(\ref{stab}) and (\ref{stat}) that the necessary and
sufficient condition for BEC to have a metastable state is $|\gamma| <
\gamma_c$.

Let $u_i(t)\equiv d_i(t)-d_{\rm st}$ be the deviations of $d_i(t)$ from
its stationary point at time $t$.
Small oscillations of these quantities are governed by
\begin{equation} \label{eqofm}
\ddot{u}_i(t) = -\sum_{j=x,y,z}
\frac{\partial^2V_{\rm eff}}{\partial d_i\partial d_j}
\Biggr|_{d = d_{\rm st}} u_j(t).
\end{equation}
Substituting $u_i(t)=u_i(0)e^{-i\omega t}$ into this yields an
eigenvalue matrix equation:
\begin{equation}
\left[ \begin{array}{ccc}
\omega^2-d_{\rm st}^{-4}-3 & d_{\rm st}^{-4}-1 & d_{\rm st}^{-4}-1 \\
d_{\rm st}^{-4}-1 & \omega^2-d_{\rm st}^{-4}-3 & d_{\rm st}^{-4}-1 \\
d_{\rm st}^{-4}-1 & d_{\rm st}^{-4}-1 & \omega^2-d_{\rm st}^{-4}-3
\end{array} \right]
\left[ \begin{array}{ccc}
u_x \\ u_y \\ u_z
\end{array} \right] = 0.
\end{equation}
For this equation to have nontrivial solutions, we should have
\begin{eqnarray} \label{det}
& & {\rm det}\! \left(\omega^2\delta_{ij}-
\frac{\partial^2V_{\rm eff}}{\partial d_i\partial d_j}
\right)\Biggr|_{d = d_{\rm st}}
\nonumber \\
& & =(\omega^2+d_{\rm st}^{-4}-5)(\omega^2-2d_{\rm st}^{-4}-2)^2=0.
\end{eqnarray}
Hence, we have the frequency of the monopole mode $\omega_{\rm M}$ and
that of the doubly degenerate quadrupole mode $\omega_{\rm Q}$
as~\cite{Stringari,Garcia}
\begin{eqnarray}
\omega_{\rm M}&=&\sqrt{5-d_{\rm st}^{-4}},    \label{mono} \\
\omega_{\rm Q}&=&\sqrt{2(1+d_{\rm st}^{-4})}. \label{quad}
\end{eqnarray}
The eigenvector corresponding to the monopole mode is given by
$(u_x,u_y,u_z)=(1,1,1)$, and therefore 
$d_x(t)=d_y(t)=d_z(t)\equiv r(t)$
always hold. Substituting this into Eq.~(\ref{act2}) yields an
effective action for the monopole mode,
\begin{equation}\label{act3}
S_{\rm M} = \frac{N \hbar}{4}\int dt\left[
3\dot{r}^2 - f(r) + 3\dot{r} \right],
\end{equation}
where
\begin{equation}\label{f}
f(r)=3r^{-2}+3r^2+\gamma r^{-3}
\end{equation}
may be viewed as an effective potential for $r(t)$.

\begin{figure}[t]
\begin{center}
\leavevmode\epsfxsize=78mm \epsfbox{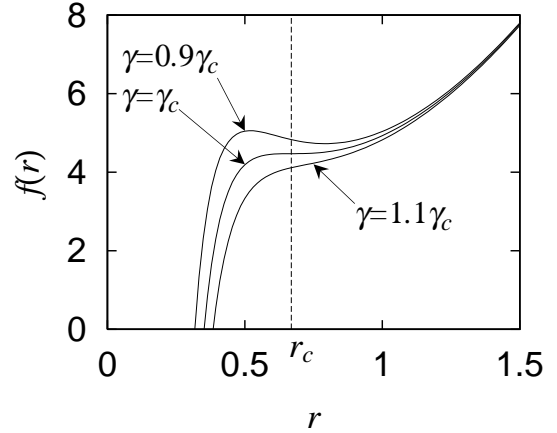}
\end{center}
\caption{
The effective potential $f(r) = 3 r^2 + 3 r^{-2} + \gamma r^{-3}$ for the
width $r$ of BEC with attractive interactions for $|\gamma| = \gamma_c = 8
\cdot 5^{-5/4}$, $|\gamma| = 0.9 \gamma_c$, and $|\gamma| = 1.1 \gamma_c$.
The dashed line shows the inflection point $r_c$ at $|\gamma| = \gamma_c$.
}
\label{f:potential}
\end{figure}

The forms of the function $f(r)$ in the vicinity of $|\gamma| = \gamma_c$
are illustrated in Fig.~\ref{f:potential}~\cite{Garcia,Sackett97}.
The potential $f(r)$ is a monotonically increasing function when 
$|\gamma|>\gamma_c$ and has a local minimum when $|\gamma|<\gamma_c$,
indicating that $\gamma_c$ is the critical strength of the interaction
above which the condensate collapses upon itself.
In other words, when the number of atoms exceeds the critical value
\begin{equation}
N_c \equiv \frac{2\sqrt{2\pi} d_{\rm st}}{5^{5/4}|a|},
\label{N_c}
\end{equation}
the parameter $r(t)$
slides down the potential in time to $r(t) \rightarrow 0$, and the central
density of the wave function grows unlimitedly.
In contrast, when the number of atoms is smaller than $N_c$, $f(r)$ has a
local minimum at which a metastable BEC is formed.

For $|\gamma|$ slightly below $\gamma_c$, we set
$|\gamma| = \gamma_c - \delta\gamma$ and $d_{\rm st} = r_c + \delta d$, and
expand Eqs.~(\ref{stat}), (\ref{mono}), and (\ref{quad}) in small
quantities $\delta\gamma$ and $\delta d$.
We then obtain
\begin{eqnarray}
\frac{\delta d}{r_c} & = & \sqrt{\frac{2}{5} \left( 1 - \frac{N}{N_c}
\right)}, \label{deltar} \\
\omega_{\rm M} & = & \sqrt{20 \frac{\delta d}{r_c}},
\label{omega_m} \\
\omega_{\rm Q} & = & \sqrt{12 - 40\frac{\delta d}{r_c}}.
\label{omega_q}
\end{eqnarray}
Substituting Eq.~(\ref{deltar}) into Eqs.~(\ref{omega_m}) and (\ref{omega_q}),
we obtain
\begin{eqnarray}
\omega_{\rm M} & = & 160^{1/4} \left( 1 - \frac{N}{N_c} \right)^{1/4}, 
\label{power} \\
\omega_{\rm Q} & = & \sqrt{12 - 8 \sqrt{10 \left( 1 - \frac{N}{N_c}
\right)}}.
\end{eqnarray}
The result (\ref{power}) agrees with that of Ref.~\cite{Ueda98} up to the
numerical coefficient.

\begin{figure}[t]
\begin{center}
\leavevmode\epsfxsize=78mm \epsfbox{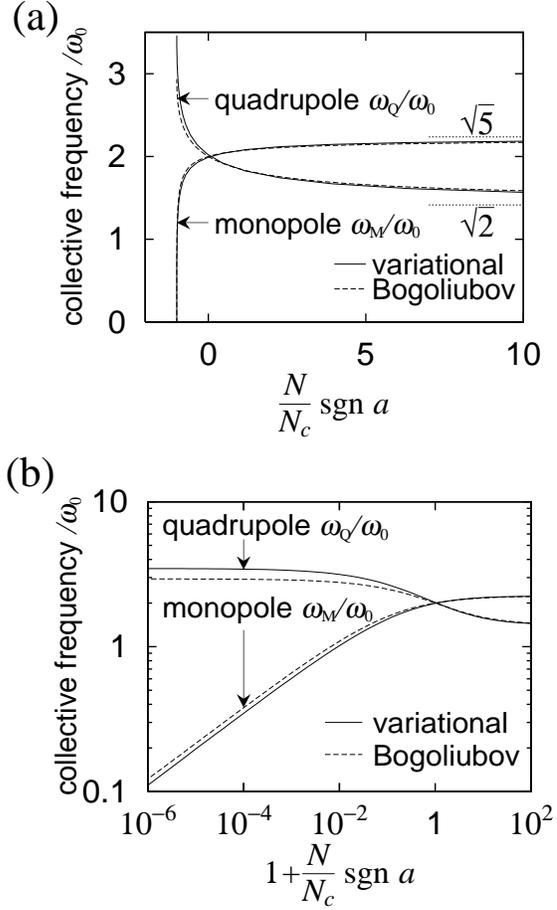}
\end{center}
\caption{
The frequencies of the monopole and quadrupole modes $\omega_{\rm M}$ and
$\omega_{\rm Q}$, normalized by the trap frequency $\omega_0$
as a function of $N / N_c \, {\rm sgn} \, a$, where ${\rm sgn} \, a = 1,
0$ and $-1$ for $a > 0$, $a = 0$, and $a < 0$, respectively.
The solid curves show the results of the variational method, and the
dashed curves show the results obtained by numerically diagonalizing the
Bogoliubov equations.
The dotted lines at $\protect\sqrt{5}$ and $\protect\sqrt{2}$ show the
Thomas-Fermi limits
of the monopole and quadrupole modes, respectively.
The curves in (b) show the same data as in (a) on a logarithmic scale to
clarify the power-law behavior of the monopole mode.
}
\label{f:omega}
\end{figure}

Figure \ref{f:omega} compares the analytic results (\ref{mono}) and
(\ref{quad}) (solid curves) with those (dashed curves) obtained by
numerically diagonalizing the Bogoliubov equations~\cite{Edwards}.
Since the Gaussian approximation overestimates the critical number of
atoms $N_c$, the numerically exact value of $N_c$ is used for the
dashed curves in Fig.~\ref{f:omega}.
In Fig.~\ref{f:omega} (a), $\sqrt{5}$ and $\sqrt{2}$ indicate the
Thomas-Fermi limits for the monopole and quadrupole modes,
respectively~\cite{Stringari}.
The numerical and analytic results are in excellent agreement for an
extensive range of $1 - N / N_c$, and the power law of $\omega_{\rm M}$
for $1 - N / N_c \ll 1$ described in Eq.~(\ref{power}) is clearly
illustrated in Fig.~\ref{f:omega} (b).

\section{Collapsing dynamics of BEC}
\label{s:4}

We assume that BEC is initially prepared in a metastable state with
$\gamma$ slightly smaller than $\gamma_c$.
We then suddenly tighten the trap potential or increase $|a|$ by a
technique such as the Feshbach resonance~\cite{Inouye,Cornish}, so that
$|\gamma| = \gamma_c + \delta \gamma$ slightly exceeds $\gamma_c$.
BEC then begins to collapse.
Although the collapsing behavior and its time scale are numerically
studied in Refs.~\cite{Garcia,Sackett98,Elef}, no analytic forms have been 
reported for them.

The effective Lagrangian for the monopole mode is obtained from 
Eq.~(\ref{act3}) as
\begin{equation} \label{L_M}
L_{\rm M} = \frac{N \hbar}{4} (3 \dot{r}^2 - 3 r^{-2} - 3 r^2 + |\gamma|
r^{-3}),
\end{equation}
and the equation of motion reads
\begin{equation} \label{rddot}
\ddot{r}(t) = r^{-3}(t) - r(t) - \frac{|\gamma|}{2} r^{-4}(t).
\end{equation}
Integrating Eq.~(\ref{rddot}) with initial conditions $r(0) = r_c$ and
$\dot r(0) = 0$ gives
\begin{equation}\label{eomr}
\frac{dr}{dt} = -\sqrt{\frac{1}{3} [f(r_c) - f(r)]}.
\end{equation}
Taking $r(t) = r_c - x(t)$ and expanding the right-hand side of
Eq.~(\ref{eomr}) up to the third power of $x(t)$, we obtain
\begin{equation} \label{eomx}
\left(\frac{dx(t)}{dt}\right)^2 =
5 \delta \gamma x(t) + 2 \cdot 5^{5/4} \delta \gamma x^2(t)
+ \frac{4\cdot5^{5/4}}{3}x^3(t).
\end{equation}
The initial stage of the collapse, where $x(t) \ll 1$, is dominated by
the first two terms on the right-hand side of Eq.~(\ref{eomx}).
Integrating Eq.~(\ref{eomx}) by keeping only these two terms, we obtain
\begin{equation} \label{tsmall}
x(t) \simeq \frac{1}{4 \cdot 5^{1/4}} \left(\cosh \sqrt{2 \cdot 5^{5/4}
\delta \gamma} t - 1 \right)
\simeq\frac{5}{4}\gamma_c\left(\frac{N}{N_c}-1\right)t^2.
\end{equation}
In contrast, the final stage of the collapse is dominated by the last term
in Eq.~(\ref{eomx}) if $\delta\gamma \ll 1$, so that we obtain
\begin{equation} \label{collapse}
x(t) \simeq \frac{3 \cdot 5^{-5/4}}{(t_{\rm collapse} - t)^2},
\end{equation}
where the constant of integration $t_{\rm collapse}$ may be interpreted 
as the time scale for BEC to collapse.
Because $t_{\rm collapse}$ is dominated by the slow initial stage of the
collapse, its evaluation requires the inclusion of the first and third
terms of Eq.~(\ref{eomx}), obtaining
\begin{eqnarray} \label{T^c}
t_{\rm collapse} & = & \int_0^\infty \frac{dx}{\sqrt{\frac{20}{3} \cdot
5^{1/4} x^3 + 5 \delta\gamma x}}
\nonumber \\
& \simeq & \frac{(3/10)^{1/4}}{4\sqrt{\pi}} \Gamma \left( \frac{1}{4}
\right)^2 \left( \frac{N}{N_c} - 1 \right)^{-1/4}
\nonumber \\
& \simeq & 1.37 \left( \frac{N}{N_c} - 1 \right)^{-1/4}.
\end{eqnarray}
It is interesting to note that this collapse time obeys the same power
with respect to $|1 - N / N_c|$ as that of $\omega_{\rm M}^{-1}$ for $N$
just below $N_c$ (see Eq.~(\ref{power})).

\begin{figure}[t]
\begin{center}
\leavevmode\epsfxsize=78mm \epsfbox{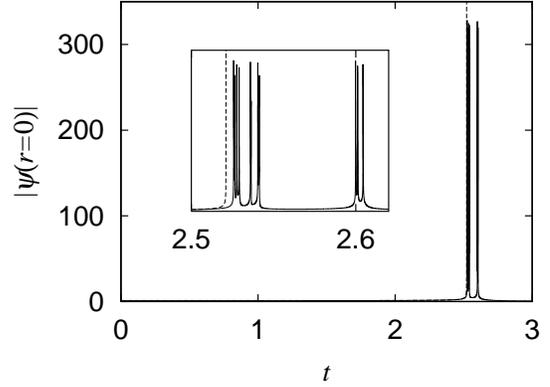}
\end{center}
\caption{
Time evolution of the peak value $|\psi(r = 0, t)|$ of the wave function.
The solid curve shows a case with two-body dipolar loss and with
three-body recombination loss (Eq.~(\protect\ref{GPloss})), and the dashed
curve without these losses (Eq.~(\protect\ref{GP})).
We first prepare BEC in a metastable state slightly below the critical
point, and at $t = 0$, increase $N$ so that $N / N_c - 1 = 10^{-2}$. 
The inset shows a blow-up view of the intermittent implosion.
}
\label{f:implosion}
\end{figure}

To check the validity of these analytic results, we numerically integrate
the time-dependent GP equation using the finite-difference method with the 
Crank-Nicholson scheme~\cite{Ruprecht}.
Figure~\ref{f:implosion} shows the time evolution of the peak height of
the wave function $|\psi(r = 0, t)|$, where we prepare an initial
metastable state for $|\gamma| / \gamma_c > 1 - 10^{-7}$, suddenly
increase $|\gamma|$ to $|\gamma| / \gamma_c = 1 + 10^{-2}$ at $t = 0$, and 
then let the state evolve in time according to the GP equation (\ref{GP})
without atomic loss or to the GP equation~(\ref{GPloss}) with the atomic
loss.
The dashed curve is obtained by solving Eq.~(\ref{GP}), and the solid
curve is obtained by solving Eq.(\ref{GPloss}) which includes the
atomic-loss processes due to two-body dipolar loss and three-body
recombination loss.
In the initial stage of the collapse, the peak density grows slowly, and
at $t \simeq 2.52$ the rapid implosion breaks out, where its blowup is
shown in the inset.
When the atomic loss is not included, the peak density increases to
infinity (dashed curve).

\begin{figure}[t]
\begin{center}
\leavevmode\epsfxsize=78mm \epsfbox{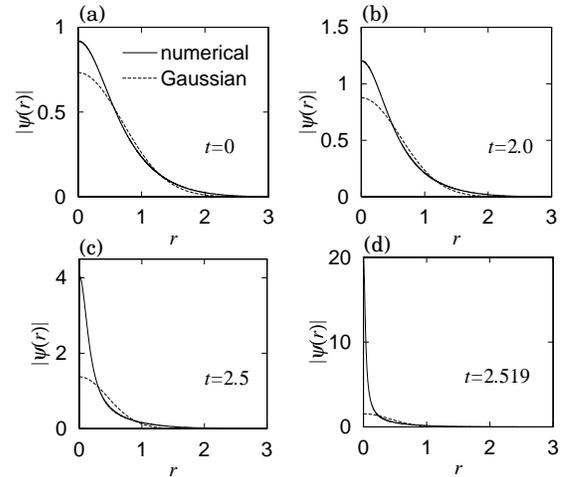}
\end{center}
\caption{
The solid curves show the profiles of the wave functions $|\psi(r, t)|$ at 
$t = 0, 2, 2.5$, and $2.519$, obtained by numerically solving the
Gross-Pitaevskii equation (\protect\ref{GP}) for the same conditions as in 
Fig.~\protect\ref{f:implosion}.
The dashed curves show Gaussian functions having the same $\langle
\hat r \rangle$ as each wave function (for the definition of $\langle \hat 
r \rangle$, see the text).
}
\label{f:wf}
\end{figure}

Figure~\ref{f:wf} shows the profiles of the wave functions $|\psi(r, t)|$
obtained by numerically solving the GP equation without loss
processes~(\ref{GP}) (solid curves) and the corresponding Gaussian
functions that has the same width $\langle \hat r \rangle$ (dashed
curves).
Even at $t = 0$, the wave function deviates from the Gaussian function,
particularly near the center of BEC.
This deviation is the origin of the 17 \% discrepancy in $N_c$ between the
numerically exact value and that obtained using the Gaussian trial wave
function.
During the initial stage of a gradual increase in the peak density, the
deviation between the numerically obtained and Gaussian wave functions
does not grow so much (see Fig.~\ref{f:wf} (b)).
At and after the outbreak of the implosion (Figs.~\ref{f:wf} (c) and (d)), 
however, the deviation becomes significant, and the Gaussian approximation
apparently breaks down, as the implosion occurs in the small region at the
center.
The implosion occurs suddenly, so we may define $t_{\rm implosion}$ as the 
time at which the rapid implosion occurs.

\begin{figure}[t]
\begin{center}
\leavevmode\epsfxsize=78mm \epsfbox{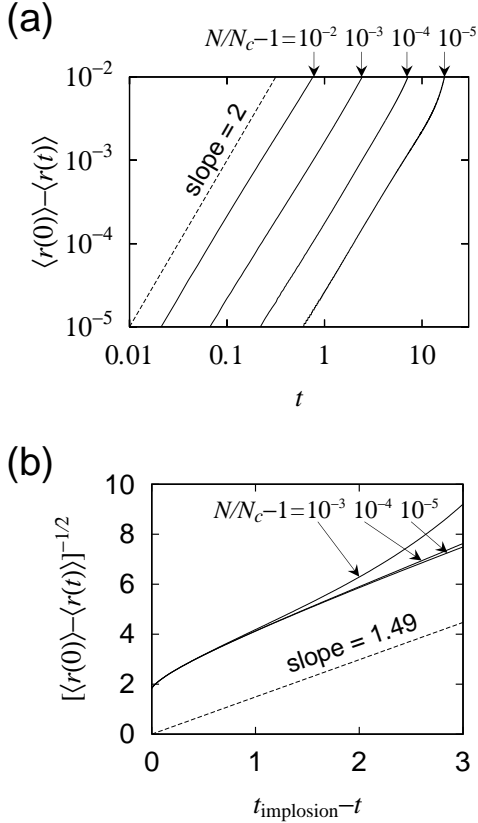}
\end{center}
\caption{
(a) The time development of $\langle \hat r(0) \rangle - \langle \hat r(t)
\rangle$ obtained by numerically solving the Gross-Pitaevskii equation
(\protect\ref{GP}) for $N / N_c - 1 = 10^{-2}$, $10^{-3}$, $10^{-4}$, and
$10^{-5}$.
BEC is initially prepared in a metastable state for $N$ just below $N_c$,
and $N$ is suddenly increased above $N_c$.
(b) The time development of $[\langle \hat r(0) \rangle - \langle \hat
r(t) \rangle ]^{-1/2}$ obtained by numerically solving the
Gross-Pitaevskii equation (\protect\ref{GP}) for $N / N_c - 1 = 10^{-3}$,
$10^{-4}$, and $10^{-5}$, as functions of $t_{\rm implosion} - t$, where
$t_{\rm implosion}$ is the time at which the rapid implosion occurs.
The dashed line shows the slope of $(3 \cdot 5^{-5/4} 2 /
\protect\sqrt{\pi})^{-1/2} \simeq 1.49$ for reference.
}
\label{f:dynamics}
\end{figure}

Until the rapid implosion takes place, the Gaussian trial wave function
fairly well describes the collapsing dynamics.
Figure \ref{f:dynamics} shows the time evolution of $\langle \hat r(0)
\rangle - \langle \hat r(t) \rangle$ at the initial and final stages
of the collapse, which is obtained by numerically solving the lossless GP
equation~(\ref{GP}).
The expectation value $\langle \hat r(t) \rangle$ is defined as
\begin{equation}
\langle \hat r(t) \rangle \equiv \int r |\psi_{\rm var}({\bf r}, t)|^2
d{\bf r} = \frac{2}{\sqrt{\pi}} r(t),
\end{equation}
where $\psi_{\rm var}({\bf r}, t)$ denotes the Gaussian variational wave
function (\ref{Gtrial}) and $r(t)$ is the variational parameter, and hence 
$\langle \hat r(0) \rangle - \langle \hat r(t) \rangle$ corresponds to $2
x(t) / \sqrt{\pi}$, where $x(t) \equiv r_c - r(t)$.
The initial changes in the width of the wave functions, shown in
Fig.~\ref{f:dynamics} (a), fit the square of $t$ well, and the prefactors
are proportional to $N / N_c - 1$, in agreement with Eq.~(\ref{tsmall}).
Figure \ref{f:dynamics} (b) shows $[\langle \hat r(0) \rangle - \langle
\hat r(t) \rangle ]^{-1/2}$ at the final stage of the collapse.
The curves become linear as they approach $t = t_{\rm implosion}$, with
the slope being close to $(3 \cdot 5^{-5/4} \cdot 2 / \sqrt{\pi})^{-1/2}
\simeq 1.49$, in agreement with the inverse-square behavior in
Eq.~(\ref{collapse}).
The curves end at a finite value $[\langle \hat r(0) \rangle - \langle
\hat r(t) \rangle ]^{-1/2} \simeq 2$ at $t = t_{\rm implosion}$, since the
rapid implosion occurs at this time.

\begin{figure}[t]
\begin{center}
\leavevmode\epsfxsize=78mm \epsfbox{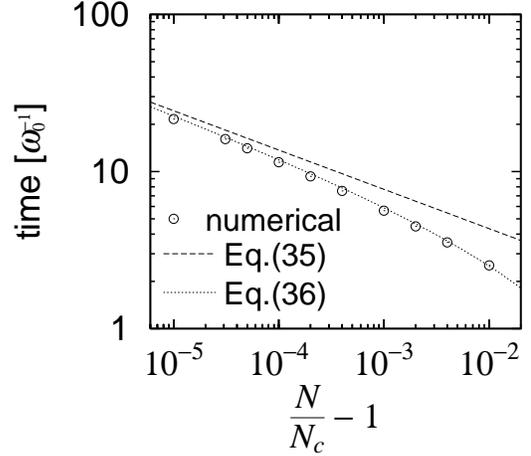}
\end{center}
\caption{
The time it takes BEC to collapse as a function of $N / N_c - 1$ for $N$
above $N_c$.
The open circles represent times at which the rapid implosions occur,
all of which are obtained by numerically solving the Gross-Pitaevskii
equation (\protect\ref{GP}).
The dashed line shows Eq.~(\protect\ref{T^c}), and the dotted curve shows
Eq.~(\protect\ref{tc}) with $\Delta t = 1.83$.
}
\label{f:collapset}
\end{figure}

The time $t_{\rm implosion}$ is determined by numerically solving the GP
equation (\ref{GP}), while $t_{\rm collapse}$, given in Eq.~(\ref{T^c}),
is analytically derived by using the Gaussian approximation and the
small-deviation approximation (\ref{eomx}).
Although these approximations break down just before the implosion,
$t_{\rm collapse}$ should still give the time scale of the collapsing 
dynamics, for the time it take BEC to collapse is dominated by the slow
initial stage, where the approximations are valid.
Quantitatively, however, $t_{\rm collapse}$ overestimates the correct
implosion time $t_{\rm implosion}$ because Eq.~(\ref{eomx}) neglects
higher-order terms that accelerate the implosion at the final stage of the 
collapse.
Figure~\ref{f:collapset} shows $t_{\rm collapse}$ (dashed line) and
numerically obtained $t_{\rm implosion}$ (open circles).
Since the difference between $t_{\rm implosion}$ and $t_{\rm collapse}$
comes from the final stage of the collapse and we consider the case in
which the initial numbers of atoms are almost the same, their relation
may be given by
\begin{equation} \label{tc}
t_{\rm implosion} = t_{\rm collapse} - \Delta t,
\end{equation}
where $\Delta t$ is a constant that should be determined from a numerical
analysis.
The dotted curve in Fig.~\ref{f:collapset} represents Eq.~(\ref{tc}) with
$\Delta t = 1.83$, which is in excellent agreement with the numerically
obtained exact $t_{\rm implosion}$.
For small $N / N_c - 1$, $t_{\rm implosion}$ asymptotically approaches
$t_{\rm collapse}$ (dashed line), which is governed by the power law
$\propto (N / N_c - 1)^{-1/4}$.

\section{Intermittent implosion and explosion}
\label{s:5}

If the atomic loss is not included in the GP equation, the peak density
grows to infinity, as shown in Fig.~\ref{f:implosion} (dashed curve).
In actual experiments, when the density becomes very high, the atomic
loss due to inelastic collisions becomes important.
We, therefore, employ the GP equation with loss processes~\cite{Kagan98}
as
\begin{equation} \label{GPloss}
i \frac{\partial}{\partial t} \psi = -\frac{1}{2} \nabla^2 \psi +
\frac{r^2}{2} \psi + g |\psi|^2 \psi - \frac{i}{2} \left( \frac{L_2}{2}
|\psi|^2 + \frac{L_3}{6} |\psi|^4 \right) \psi,
\end{equation}
where $L_2$ and $L_3$ denote the two-body dipolar and three-body
recombination loss-rate coefficients, respectively.
The two-body (three-body) loss-rate coefficient must be divided by two
(six) because of Bose statistics~\cite{factor}.
In Eq.~(\ref{GPloss}), $L_2$ and $L_3$ are made dimensionless by
multiplying $N / (\omega_0 d_0^3)$ and $N^2 / (\omega_0 d_0^6)$,
respectively.
The atoms and molecules produced by inelastic collisions are assumed to
escape from the trap without affecting the condensate.

The solid curve in Fig.~\ref{f:implosion} shows the time evolution of the
peak height of the wave function $|\psi(r = 0, t)|$ following
Eq.~(\ref{GPloss}), where we assume a ${}^7{\rm Li}$ condensate with $N =
1260$, $\omega_0 = 2 \pi \times 144.5$ Hz~\cite{Sackett99}, $L_2 = 1.05
\times 10^{-14}$ ${\rm cm}^3 / {\rm s}$~\cite{Gerton}, and $L_3 = 2.6
\times 10^{-28}$ ${\rm cm}^6 / {\rm s}$~\cite{Moer}.
The initial condition is the same as the lossless case (dashed curve).
During the gradual increase in the peak density, a few atoms are lost,
which slightly delays the time of implosion.
When the implosion begins, the role of the atomic loss becomes
significant.
The implosion stops at a certain density, showing pulse-like behavior
(see the inset of Fig.~\ref{f:implosion}), and occurs several times
intermittently in rapid sequence~\cite{Saito}.
This behavior may be qualitatively explained as follows.
When the rapid implosion occurs and the peak density satisfies
$g|\psi|^2 \sim L_3 |\psi|^4 / 12$, i.e., $|\psi|^2 \sim 12 g / L_3$, the
collisional loss rate begins to surmount the accumulation rate of atoms
at the center.
When the atoms at the peak are suddenly removed due to atomic loss, the
attractive force within a small spatial region is weakened, and the atoms
begin to explode rather than implode due to the zero-point kinetic
pressure.
When the inward flow outside the region of implosion is sufficient to
replenish the peak density, the subsequent implosion occurs.
The intermittent implosions thus occur as the result of a competition
between the loss of atoms and their accumulation due to the attractive
interaction; therefore, these implosions should be distinguished from
other oscillatory behaviors due to various mechanisms such as
collapse-and-growth cycles~\cite{Sackett97,Sackett98,Kagan98} and the
small collapses described in Ref.~\cite{Kagan98}.
In the case of ${}^7{\rm Li}$, the peak density is estimated to be
$|\psi|^2 \sim 12 g / L_3 \simeq 400^2$, which qualitatively agrees with
the solid curve in Fig.~\ref{f:implosion}.
We should note here that the peak density $|\psi|^2 \sim 48 \pi \hbar a /
(m L_3)$, where the dimension is restored, is independent of the number of
atoms $N$ and the trap frequency $\omega_0$, indicating that the
implosions are a local phenomenon depending only on the nature of the
atoms themselves.

\begin{figure}[t]
\begin{center}
\leavevmode\epsfxsize=78mm \epsfbox{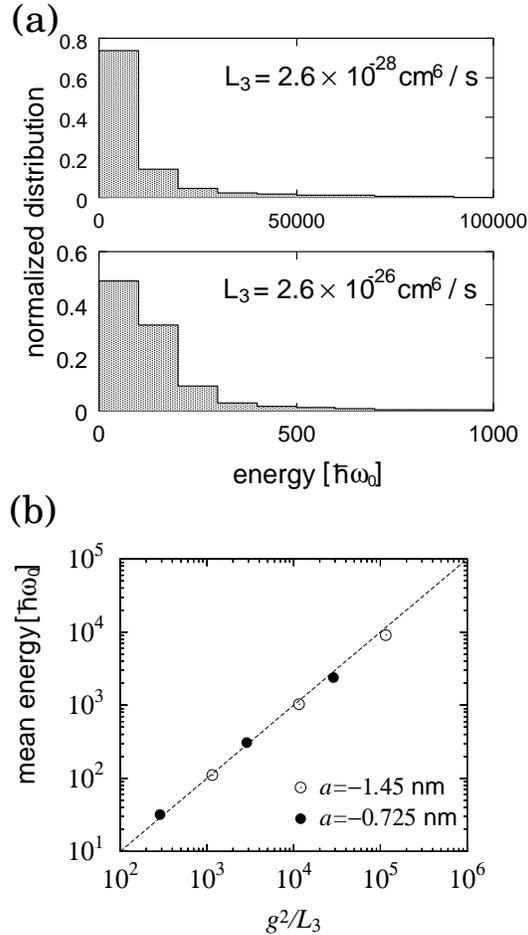}
\end{center}
\caption{
(a) Normalized distributions of the energy of an atom scattered by the
explosion.
The same parameters and the initial conditions as those in the solid curve 
in Fig.~\protect\ref{f:implosion} are used in the upper distribution, and
the three-body recombination rate is multiplied by 100 in the lower one.
(b) The mean energies of an scattered atom.
The rightmost open circle is obtained using the parameters of ${}^7{\rm
Li}$, and the other open circles are obtained by increasing the three-body
recombination rate by 10 and 100.
In the solid circles, the s-wave scattering length is halved.
}
\label{f:explosion}
\end{figure}

The explosion following an implosion of atoms is predicted in
Refs.~\cite{Kagan98,Saito} and has been experimentally observed as a
broadly scattered ``hot'' atom cloud of about 100 nK~\cite{Cornish}.
From the above discussion, the energy scale of the explosion might be
estimated as $g|\psi|^2 \sim 12 g^2 / L_3$.
This energy is, however, the highest one that only a small number of atoms
at the center of BEC may acquire.
In fact, the atoms scattered by the explosion have a broad momentum
distribution.
Figure~\ref{f:explosion} (a) shows a normalized energy distribution of
atoms scattered by the explosion, where the parameters of ${}^7{\rm Li}$
are used; the lower panel of Fig.~\ref{f:explosion} (a) shows the same
distribution with $L_3$ is multiplied by 100.
These energy distributions are obtained from the wave functions at $t =
t_{\rm implosion} + \pi / 2$, at which the ejected part of BEC spreads
maximally and are therefore easy to be discerned from the remnant part of
BEC that remains at the center.
The energy per atom at position ${\bf r}$ is given by
\begin{equation}
\psi^*({\bf r}) \left( -\frac{1}{2} \nabla^2 + \frac{r^2}{2} \right)
\psi({\bf r}) / |\psi({\bf r})|^2,
\end{equation}
where the interaction energy is omitted since the density of the
widespread atoms are low enough.
In obtaining the energy distributions and the mean energies, the
remnant part of the condensate at the center is excluded.
While the energy scale is different, the two distributions in
Fig.~\ref{f:explosion} (a) have similar shapes.
The number of atoms scattered by the explosion is 20 $\sim$ 40 \% of the
total number of atoms.
The dependence of this fraction on the parameters $g$ and $L_3$ is
complicated, since it depends on the number of times and the detailed
structure of the intermittent implosions that are governed by highly
nonlinear dynamics.

Figure~\ref{f:explosion} (b) shows the mean energies of the atoms
scattered by the explosion as a function of $g^2 / L_3$.
The open circles are obtained using the value of $g$ of ${}^7{\rm Li}$
with $\omega_0 = 2 \pi \times 144.5$ Hz and the three-body recombination
rates $L_3$, $10 L_3$, and $100 L_3$.
In order to see the dependence on $g$, the value of $g$ is halved in the
solid circles.
The open and solid circles have almost the same dependence on $g^2 / L_3$,
suggesting that the explosion energy is determined by the ratio $g^2 /
L_3$, and not by $g$ and $L_3$ separately.
The mean energy is roughly given by $\sim 0.1 g^2 / L_3$, or
restoring the dimension by
\begin{equation} \label{meanE}
\sim 0.1 \times 16 \pi^2 \frac{\hbar^3 a^2}{m^2 L_3}.
\end{equation}
We note that this energy is independent of the number of atoms $N$ and the 
trap frequency $\omega_0$, as with the peak density discussed above.
These results indicate that the global nature of the system is unimportant
with regard to implosion and explosion, and that only the local nature of
the atoms determines their behavior.
Using Eq.~(\ref{meanE}), the mean kinetic energy of the atoms scattered by
the explosion in the experiment at Rice~\cite{Sackett99} is estimated to
be $\simeq 80 \mu{\rm K}$.
In the experiment at JILA~\cite{Cornish}, on the other hand, the
three-body recombination rate of ${}^{85}{\rm Rb}$ is uncertain in the
vicinity of the Feshbach resonance, where it significantly depends on the
applied magnetic field~\cite{Roberts}.
The experiment of collapse is carried out by applying magnetic field of
about 170 G at which sign of the interaction changes.
Because of uncertainty in the three-body recombination rate, we estimate
it from the experimental result of the explosion energy $\simeq 100$ nK
based on Eq.~(\ref{meanE}).
Assuming $a = -1$ nm, we obtain $L_3 \simeq 6 \times 10^{-28}$ ${\rm cm}^6
/ {\rm s}$, which is consistent with the value of $L_3$ around 170
G~\cite{Roberts}.
The above-described method thus provides an independent way to determine
the three-body recombination rate in addition to the current
methods~\cite{Burt,Kurn,Roberts}.

\section{Conclusions}
\label{s:conclusion}

We have analytically and numerically studied the dynamics of BEC with
attractive interactions below and above the critical number of atoms
$N_c$.
We have shown that our analytic approach successfully describes the
collective oscillations below $N_c$ as well as the collapsing dynamics of
BEC above $N_c$ until the rapid implosion takes place.
We have found that the time scales of the
dynamics follow the same power law on either side of the critical point.
Just below $N_c$, the collective frequency of the monopole mode is
proportional to $(1 - N / N_c)^{1/4}$, and just above $N_c$, the time it
takes BEC to collapse is proportional to $(N / N_c - 1)^{-1 / 4}$.

When the atomic loss due to inelastic collisions is taken into account,
the implosion occurs not only once but several times intermittently, and
the implosion is then converted to an explosion.
We have found that the mean energy of an atom scattered by the explosion
is given by Eq.~(\ref{meanE}), which is independent of the number of atoms
and the trap frequency.

\section*{ACKNOWLEDGMENTS}

This work was supported by a Grant-in-Aid for Scientific Research (Grant
No. 11216204) by the Ministry of Education, Science, Sports, and Culture
of Japan, and by the Toray Science Foundation.

\end{document}